\documentstyle[aps,preprint]{revtex}

\def \grd {\mbox {\boldmath $\nabla$} }
\def \bphi {\mbox {\boldmath $\phi$} }
\def \bpsi {\mbox {\boldmath $\psi$} }
\def \bdelta {\mbox {\boldmath $\delta$} }
\def \bsigma {\mbox {\boldmath $\sigma$} }

\begin{document}
\draft
\preprint{}
\title  {Hydrodynamic fluctuations in the Kolmogorov flow: \\ Nonlinear regime}
\author{ I. Bena$^{a}$,  F.  Baras$^{b}$ and M. Malek  Mansour$^{b}$\\}

\address{(a)  Limburgs Universitair Centrum\\ B-3590 Diepenbeek, Belgium \\
(b) Centre for Nonlinear Phenomena and Complex Systems\\ Universit\'e Libre de
Bruxelles, \\ Campus Plaine, C.P. 231 \\ B-1050 Brussels, Belgium \\}

\date{\today}

\maketitle

\begin{abstract} \\In a previous paper \cite{malek1} the
statistical properties of the linearized Kolmogorov flow have been studied,
using the formalism of fluctuating hydrodynamics.  In this paper the nonlinear
regime is considered, with emphasis on the statistical properties of the flow
near the first instability. The normal form amplitude equation is derived
for the case of an incompressible fluid and the velocity field is constructed
explicitly above (but closed to) the instability.  The relative simplicity
of this flow allows one to analyze the compressible case as well.  Using a
perturbative technique, it is shown that close to the instability threshold
the stochastic dynamics of the system is governed by two coupled non linear
Langevin equations in the Fourier space.  The solution of these equations
can be
cast into the exponential of a Landau Ginzburg functional, which proves to be
identical to the one obtained for the case of the incompressible fluid.  The
theoretical predictions are confirmed by numerical simulations of the nonlinear
fluctuating hydrodynamic equations.

\end{abstract}
\

\pacs{05.40+j, 05.90+m, 47.20.-k, 42.70.Ft, 47.40.Dc}

\section{Introduction}
\label{sec:intro}

A central issue in nonequilibrium statistical physics is the role of
fluctuations in the onset of hydrodynamical instabilities.  From a theoretical
point of view one generally relies on the Landau-Lifshitz fluctuating
hydrodynamics \cite{landau}, mainly because of its relative simplicity as
compared to more fundamental approaches
\cite{others1,dufty}.  Fluctuating hydrodynamics has been used by various
authors
to study the statistical properties of simple fluids subjected to
nonequilibrium
constraints, such as temperature gradient
\cite{others1,others2,mazur} or shear
\cite{dufty,machta} (for a review, see ref. \cite{schmitz}). Light
scattering results, obtained for systems under temperature gradient, have shown
quantitative agreement with theoretical predictions \cite{law}. Quantitative
agreement has also been demonstrated with results based on particle
simulations, both for systems under temperature gradient
\cite{malek,jpb} and shear \cite{garcia}.

Ordinarily the macroscopic study of subsonic hydrodynamical instabilities is
based on the incompressibility assumption.  However, as first pointed out
by Zaitsev and
Shliomis \cite{zaitsev}, this assumption is basically inconsistent with the
very
foundations of the fluctuating hydrodynamics formalism since it imposes
fictitious correlations between the velocity components of the fluid.  On the
other hand, the compressibility of the fluid affects mostly fast sound modes,
whereas the dynamics of the system near an instability is governed by
slow dissipative modes. We may thus expect that the behavior of a fluid
evolving
near a subsonic instability threshold is practically not affected by its
compressibility.
This intuitive argument has been used by many authors who have considered
fluctuating incompressible hydrodynamic equations, or even directly the
corresponding normal form amplitude equations to which they added random noise
terms \cite{graham}.  In these approaches, the characteristics of the noise
terms cannot be related to equilibrium statistical properties of the fluid and
thus remain arbitrary.  A more satisfactory approach would be to start with the
full compressible fluctuating hydrodynamic equations.  Reducing these equations
to a final normal form amplitude equation near the instability, would lead
directly to the explicit form of the associated noise terms consistent with
such requirements as the fluctuation-dissipation theorem.  Such a procedure,
however, proves to be quite difficult mainly because of the boundary
conditions.  To our knowledge, the only attempt in this direction has been made
by Schmitz and Cohen for the case of B\'enard instability
\cite{cohen}.  Concentrating on the behavior of a small layer in the bulk,
these
authors have succeeded in deriving the linearized fluctuating equations
close to
the convective instability.  Whether this technique can be generalized to
derive
the corresponding normal form amplitude equation for the case of B\'enard
instability is not clear at present time.

Recently, we have considered the problem of hydrodynamic fluctuations in the
case of a simple flow proposed some fifty years ago by Kolmogorov \cite{obu}.
Thanks to the periodic boundary conditions associated to this  model, a
detailed analysis of the linearized fluctuating hydrodynamic equations, from
near equilibrium up to the vicinity of the first instability, could be carried
out
\cite{malek1}.  In particular, we have been able to show that in the
long time limit the flow behaves basically as if the fluid were incompressible,
regardless of the value of the Reynolds number.  The situation was
different for
the short time behavior.  We established that the
incompressibility assumption leads here to  a wrong  form of the
static correlation functions, in agreement with Zaitsev and Shliomis prediction
\cite{zaitsev}, except near the instability threshold, where our results
strongly suggest that the incompressibility assumption becomes again valid.  On
the other hand, the linearized fluctuating hydrodynamic equations are clearly
not valid close to, or beyond the instability threshold. Although extensive
numerical simulations have basically confirmed our predictions, a satisfactory
answer to this important problem requires a full nonlinear analysis of the
fluctuating Kolmogorov flow.  The present article is devoted to this
problem.

In the next section, the  Kolmogorov flow is briefly reviewed.  A
nonlinear analysis is carried out for an incompressible fluid and the
explicit form of the stream function and the associated velocity field is
derived above but close to the instability.  Section III is devoted
to the analysis of a compressible fluid.  After setting up a perturbation
scheme, we show that the solution of the problem is basically the same as the
one derived in section II for the incompressible fluid, at least close to the
instability threshold.  We then concentrate on the statistical properties
of the
flow and show that, close to the instability threshold, the dynamics of the
system is governed by a set of two nonlinear coupled Langevin equations.  Here
again, the equivalence with the incompressible case is established.
Concluding remarks and perspectives are summarized in section IV.

\section{Incompressible Kolmogorov flow}
\label{sec:II}

\par Consider an isothermal flow in a rectangular box $L_x \times L_y$ oriented
along the main axes, that is
$\left\{ 0 \le x < L_x  ,  0 \le y < L_y \right\}$.  Periodic boundary
conditions are assumed in both directions and the flow is maintained through an
external force field of the form

\begin{equation}
 {\bf F}_{ext}=F_0\,\sin{(2\,\pi\,n\,y/L_y)}\,{\bf 1}_x\,,
\end{equation} where ${\bf 1}_x$ is the unit vector in the $x$ direction. This
model represents the so-called  {\it Kolmogorov flow} and it belongs to the
wider
class of two-dimensional negative eddy viscosity flows \cite{gama}. It is
entirely characterized through the strength of the force field $F_0$, the
parameter $n$, which controls the wave number of the forcing, and the aspect
ratio
$a_r$, defined as
\begin{equation} a_r = L_x/L_y \,.
\end{equation} In the following, we will mainly concentrate on the case
$n\,=\,1$.

The fluctuating hydrodynamic equations for this model read:
\begin{equation}
\frac{\partial \,\rho}{\partial\,t} \, = \, -\grd\,\cdot\,(\rho\,{\bf   v})\,,
\label{continuity}
\end{equation}
\begin{equation}
\rho\,\frac{\partial\,{\bf   v}}{\partial\,t} \, = \, -\rho\,({\bf
v}\cdot\grd)\,{\bf v}\,-\,\grd\,p \,- \, \grd \cdot \bsigma \,+\, {\bf
F}_{ext}\,,
\label{momentum}
\end{equation} where $\rho$ is the mass density, $p$ the hydrostatic
pressure and
$\bsigma$ the {\it two dimensional} fluctuating stress tensor:
\begin{equation}
\sigma_{i, j} \, = \, - \, \eta \, \left(
\frac{\partial\,v_i}{\partial\,x_j} + \,
\frac{\partial\,v_j}{\partial\,x_i} - \, \delta_{i, j} \, \grd \cdot {\bf v}
\right) \, - \,
\zeta \, \delta_{i, j} \, \grd \cdot {\bf   v}\,+\,S_{i,\,j}\,.
\label{stress}
\end{equation} ${\bf{S}}$ is a random tensor whose elements $\{S_{i,\,j}\}$ are
Gaussian white noises with zero mean and covariances given by:
\begin{eqnarray} < S_{i, j}({\bf   r}, t) \, S_{k, \ell}({\bf   r}', t') >
& = &
2 k_B T_0 \,
\delta (t - t')  \, \delta ({\bf   r} -{\bf   r}') \left [ \eta (\delta_{i,
k}^{Kr}
\delta_{j, \ell}^{Kr} +
\delta_{i,\ell}^{Kr} \delta_{j, k}^{Kr} ) \, + \, (\zeta - \eta)
\delta_{i, j}^{Kr}
\delta_{k, \ell}^{Kr} \right ].\nonumber \\
\label{noisecor}
\end{eqnarray} For simplicity, we shall assume that the shear and bulk
viscosity
coefficients,
$\eta$ and
$\zeta$, are {\it state independent}, i.e. they are constant.

Let us first concentrate on the {\it deterministic} behavior. It can be easily
checked that at the stationary state, the pressure  and the  density  are
uniform in space ($p_{st}\,=\,p_0,\,\rho_{st}\,=\,\rho_{0}$), whereas the
velocity profile is given by:
\begin{eqnarray} {\bf v}_{st} \, =  & u_0 &\,
\sin{(2\,\pi\,y/L_y)}\,{\bf 1}_x \,,\nonumber\\ & u_0  &\, = \,
\frac{F_0\,L_y^2}{4\,\pi^2\,\eta}\,.
\label{vstat}
\end{eqnarray}  For small enough
$F_0$, this stationary flow is stable.  As we increase $F_0$, however, the flow
eventually becomes unstable giving rise to rotating convective patterns. Other
instabilities of increasing complexity may occur for larger values of $F_0$,
culminating in a chaotic like behavior similar to what is observed in turbulent
flows
\cite{lorenz,she,benzi}. In this paper we shall limit ourselves to the analysis
of the system near its first instability.

We still have to supply the momentum conservation equation (\ref{momentum})
with
an equation of state relating the pressure to the density (recall that the
system is isothermal).  In this section, we shall simply assume that the
flow is
incompressible, i.e.
\begin{equation}
\grd \, \cdot \, {\bf v}\,=\, \frac{\partial{u}}{\partial x} \, + \,
\frac{\partial{v}}{\partial y} \, = 0\,,
\label{incomp}
\end{equation} where $u$ and $v$ represent the $x$ and $y$ components of the
velocity, respectively, i.e.
${\bf v} \equiv u {\bf 1}_x + v {\bf 1}_y$.  Relation (\ref{incomp}) implies a
uniform density $\rho_0$ throughout the system for all time, if initially
so, as
well as the existence of a scalar function
$\psi (x, y)$, known as {\it stream function}, defined by the relations:
\begin{equation} u\,=\,\frac{\partial \psi}{\partial y}\,,
\,\,\,\,v\,=\,-\,\frac{\partial\psi}{\partial x}\,.
\label{streamfunction}
\end{equation} Scaling lengths by $L_y$, velocity by $u_0 $  and time by
$L_y/u_0 $, the dimensionless equation for the stream function reads:
\begin{equation}
\frac{\partial(\nabla^2\,\psi)}{\partial t}\,=\,-\,\frac{\partial \psi}
{\partial y}\,\frac{\partial(\nabla^2\,\psi)}{\partial x}
\,+\,\frac{\partial\psi}{\partial x}\,\frac{\partial(\nabla^2\,\psi)} {\partial
y}\,+\,R^{-1}\,\nabla^2\,(\nabla^2\,\psi)\,+\,8\,\pi^3\,R^{-1}\,\cos{(2\,\pi\,y)
}
\label{incompsi}
\end{equation} where $R$ is the Reynolds number:
\begin{equation} R\, = \,\frac{\rho _0 \,u_0 \,L_y}{\eta}\,.
\label{defR}
\end{equation} The stationary solution of (\ref{incompsi}) reads:
\begin{equation} {\psi}_{st}\,=\,-\,\frac{1}{2\,\pi}\,\cos{(2\,\pi\,y)}\,.
\end{equation} Setting $\psi = \psi_{st} + \delta \psi$, and linearizing
(\ref{incompsi}) around $\psi_{st}$, one gets
\begin{equation}
\frac{\partial (\nabla^2\,\delta\psi)}{\partial
t}\,=\,-\,\sin{(2\,\pi\,y)}\,\frac{\partial (\nabla^2\,\delta\psi)}{\partial x}
\,-\,4\,{\pi}^2\,\sin{(2\,\pi\,y)}\,\frac{\partial\,\delta\psi}{\partial x}
\,+\,R^{-1}\,\nabla^2\,(\nabla^2\,\delta \psi)\,.
\label{psilin}
\end{equation}
Owing to periodic boundary conditions, $\delta\psi(x, y, t)$ can
be expanded in Fourier series:
\begin{eqnarray}
\delta\psi(x,\,y,\,t)& = &
\sum_{k_x,\,k_y\,=\,-\,\infty}^{\infty}\exp{(-\,2\,\pi\,i\,k_y\,y)}
\,\exp{(-\,2\,\pi\,i\,k_x\,x/a_r)}\,\delta\psi_{k_x,\, k_y}(t)\, ,
\nonumber\\
\delta\psi_{k_x,\, k_y}(t)&=&\int_{0}^{1}dy\,\exp{(2\,\pi\,i\,k_y\,y)}
\,\frac{1}{a_r}\int_{0}^{a_r}dx\,\exp{(2\,\pi\,i\,k_x\,x/a_r)}\,
\delta\psi(x,\,y,\,t)\,.
\nonumber\\
\end{eqnarray}
Equation (\ref{psilin}) can  be then transformed to :
\begin{eqnarray}
\frac{\partial \delta\psi_{k_x,\,k_y}}{\partial t}\, = & - &
4\,{\pi}^2\,R^{-1}\,({\tilde {k}_x}^2\,+\,{k_y}^2)
\delta\psi_{k_x,\, k_y}
\nonumber\\
 & + & \pi\,\tilde{k}_x\,
 \left[\delta\psi_{k_x,\,k_y\,+\,1}\,-\,\delta\psi_{k_x,\,k_y\,-\,1}\right]
\nonumber\\ & + & \,2\,\pi\,\frac{\tilde {k}_x\,k_y}{{\tilde
{k}_x}^2\,+\,{k_y}^2}\,
\left[\delta\psi_{k_x,\,k_y\,+\,1}\,+\,\delta\psi_{k_x,\,k_y\,-\,1}\right]\,,
\label{fourlin}
\end{eqnarray}  where we have set
\begin{equation}
\tilde {k}_x = k_x/a_r\,.
\end{equation}

In its general form, the analysis of this equation proves to be quite difficult
\cite{sinai}.  On the other hand, if $\psi_{st}$ is stable then, in the
long time
limit, the evolution of the system will be  mainly governed by long wavelength
modes. Accordingly, we start our analysis by considering only the modes
$k_y\,=\,0\,,\pm 1$, i.e. we assume that $\delta\psi(k_x\,,k_y\,t)
\approx 0$ for $|k_y|\,\geq 2$ \cite{green}. Defining the vector $\bdelta
\bpsi_{k_x}
\equiv
\left(
\delta\psi_{k_x ,\,0} \, ,
\delta\psi_{k_x ,\,1} \, ,
\delta\psi_{k_x ,\,-1} \right)$ , the equation (\ref{fourlin}) can  be written
in the following matricial form :
\begin{equation}
\frac{\partial \bdelta \bpsi_{k_x}(t)}{\partial t}\,=\, {\bf A}
\,\cdot\,\bdelta \bpsi_{k_x}(t)\,,
\label{psilink}
\end{equation} with
\begin{eqnarray} {\bf A} & = &
\left(
\begin{array}{ccc} -\,4\,\pi^2\,R^{-1}\,{\tilde {k}_x}^2  &
\, \, \, \,\, \pi\,\tilde {k}_x & -\,\pi\,\tilde {k}_x \\
\, \pi\,\tilde {k}_x \, (1 - {\tilde {k}_x}^2 ) / (1 + {\tilde {k}_x}^2 ) \, &
\,\,\,-\,4 \pi^2\,R^{-1} (1\,+\,{\tilde {k}_x}^2) & 0 \\ -\,\pi\,\tilde
{k}_x \,
(1 - {\tilde {k}_x}^2 ) / (1 + {\tilde {k}_x}^2 ) \,  & \,\,\, \, 0  &
\,-\,4 \pi^2\,R^{-1} (1\,+\,{\tilde {k}_x}^2)
\end{array}
\right)
\end{eqnarray}

We first note that the matrix ${\bf A}$ is diagonal for $k_x = 0$ so that the
solution of eq. (\ref{psilink})  simply reduces to:
\begin{equation}
\delta \psi_{0,\,1}(t) \, \sim \, \delta \psi_{0,\, - 1}(t) \, \sim
\, \exp (- \, 4
\pi^2 R^{-1} \, t) .
\end{equation} Furthermore, by definition of the stream function, eq.
(\ref{streamfunction}),
$\psi_{0,\,0}(t) = 0, \, \forall \, t$.   We thus concentrate on the case
$k_x \ne 0$, looking
for a similarity transformation ${\bf T}
\cdot {\bf A} \cdot {\bf T}^{-1}$ which diagonalizes the matrix ${\bf A}$.
After
some algebra, one finds
\begin{equation} {\bf T}\,=\,
\left(
\begin{array}{ccc} (\lambda_1 - \lambda_3)  /  \pi\, \tilde {k}_x &
\, \,
\, \, 1 & \, \, \,
\, -1\\ (\lambda_2 - \lambda_3)  /  \pi\, \tilde {k}_x & \, \, \, \, 1 & \,
\, \,
\, -1\\ 0 & \, \, \, \, 1 & \, \, \, \, \,1
\end{array}
\right)\,,
\label{Tmatrice}
\end{equation} where $\left \{ \lambda_i \right \}$ are the eigenvalues of
${\bf A}$:
\begin{eqnarray}
\lambda_1 \,&=&\,-\,2\,\pi^2\,R^{-1}\,(1\,+\,2\,{\tilde
{k}_x}^2)\,+\,\pi\,\sqrt{2\,{\tilde {k}_x}^2\, (1-{\tilde {k}_x}^2) / (1
+{\tilde {k}_x}^2)
\,+\,4\,\pi^2\,R^{-2}} \,,\nonumber
\\
\lambda_2 \,&=&\,-\,2\,\pi^2\,R^{-1}\,(1\,+\,2\,{\tilde
{k}_x}^2)\,-\,\pi\,\sqrt{2\,{\tilde {k}_x}^2\, (1-{\tilde {k}_x}^2) / (1
+{\tilde {k}_x}^2)
\,+\,4\,\pi^2\,R^{-2}} \,,\nonumber
\\
\lambda_3 \,&=&\,-\,4\,\pi^2\,R^{-1}\,(1\,+\,{\tilde {k}_x}^2)\,.
\label{vpropre}
\end{eqnarray} The equation (\ref{psilink}) then becomes:
\begin{equation}
\frac{ \partial \delta \phi_i(t)}{ \partial t} \, = \, \lambda_i \,
\delta
\phi_i(t) \, ,
\,\,\,\, i \, = \, 1, 2, 3\,\,\,,
\end{equation} where
\begin{equation}
\bdelta \bphi \, = \, {\bf T} \, \cdot \,  \bdelta \bpsi
\label{Ttransform}
\end{equation}

It follows from (\ref{vpropre}) that  $\lambda_2$ and $\lambda_3$ are always
negative, whereas there exists a critical value of the Reynolds number
\begin{equation} R_c(k_x)\,=\,2\,\sqrt{2}\,\pi\,\frac{1\,+{\tilde
{k}_x}^2}{\sqrt{1\,-\,{\tilde {k}_x}^2}} \, \,  ; \, \, \, \, \, \,
\, \, 0
\, < \, {\tilde {k}_x}^2
\, < \, 1
\label{Rc}
\end{equation} for which $\lambda_1$ vanishes, thus indicating the limit of
stability of the corresponding mode \cite{nepo}.  Clearly $R_c$ is an
increasing
function of $|k_x|$, so that the first modes to become unstable correspond to
$|k_x| = 1$, provided the aspect ratio
$a_r > 1$. As $a_r\,\rightarrow\, 1$, $R_c\,\rightarrow\,\infty$, indicating
that no instability can develop for perturbations of the same spatial
periodicity
as the applied force \cite{machioro}.  In the following, we shall therefore
concentrate mainly on the case $a_r > 1$.

For $a_r = 2$,  relation (\ref{Rc}) predicts a critical Reynolds number of $R_c
\approx 12.8255$.  Analytical calculations can still be handled when the modes
$k_y = \pm 2$ are taken into account as well, and lead to
\begin{equation} R_c^{(5)}(k_x)\,=\,R_c(k_x)\,\left[1 \, + \frac{{\tilde
{k}_x}^4 \, ({\tilde {k}_x}^2 + 3)}{2\, ({\tilde {k}_x}^2+4)^{2}\,({\tilde
{k}_x}^2-1)}\right]^{- 1/2}
\,
\,  ; \, \, \, \, \, \,
\, \, 0
\, < \, {\tilde {k}_x}^2
\, < \, 1
\label{Rc5}
\end{equation} For $a_r = 2$, one finds a critical Reynolds number of
$R_c^{(5)}
\approx 12.8738$, so that the discrepancy remains below
$0.4 \%$.  Numerical evaluation of
$R_c$ performed with a total amount of $103$ modes shows no further significant
discrepancy.  We thus conclude that one can rely reasonably well on a "3-modes
approximation theory" (that is
$\delta\psi_{k_x,\,k_y}(t) \approx 0$ for $|k_y|\,\ge 2$).  It remains to check
whether this approximation leads to the correct velocity field beyond the
instability. To this end we need to work out the explicit form of the stream
function.

The calculations are tedious and quite lengthy, so that here we only report the
basic steps. We start with the full nonlinear evolution equation for $\delta
\psi =  \psi -
\psi_{st}$:
\begin{eqnarray}
\frac{\partial (\nabla^2 \delta\psi)}{\partial t} & = &
-\sin{(2\,\pi\,y)}\,\frac{\partial (\nabla^2 \delta\psi)}{\partial
x}\,-\,4\,\pi^2\,\sin{(2\,\pi\,y)}\,\frac{\partial
\delta\psi}{\partial x}\nonumber\\ & + & R^{-1}\,\nabla^2(\nabla^2\delta\psi)\,
- \,
\frac{\partial
\delta\psi}{\partial y}\,\frac{\partial (\nabla^2
\delta\psi)}{\partial x}\,+\,\frac{\partial
\delta\psi}{\partial x}\,\frac{\partial (\nabla^2
\delta\psi)}{\partial y} \,.
\label{psinlin}
\end{eqnarray} As for the linear case, we take the Fourier transform of this
equation, limiting ourselves to the first three modes  $k_y = 0, \pm 1$.
Applying
then the transformation ${\bf T}$ to the resulting equation, cfr. eq.
(\ref{Ttransform}), one obtains:
\begin{equation}
\frac{ \partial \delta \phi_i(t)}{ \partial t} \, = \, \lambda_i \,
\delta
\phi_i(t) \, + \, \Phi_i \, \, ,
\,\,\,\, i \, = \, 1, 2, 3\,\,\,,
\label{phi1}
\end{equation} where the $\Phi_i$'s are nonlinear polynomial functions of
$\delta
\phi_1, \delta \phi_2$ and $\delta \phi_3$ and their complex-conjugates. Close
to the bifurcation point ($R \approx R_c$ ,
$k_x = 1$), the mode
$\delta \phi_1$ exhibits a {\it critical slowing down} since
$\lambda_1(k_x = 1) \approx 0$.  On this slow time scale, i.e.
$t \approx O(\lambda _1^{-1})$, the fast modes $\delta \phi_2$ and
$\delta
\phi_3$ can then be considered as stationary, their time dependence arising
mainly through
$\delta \phi_1(t)$. Setting
$\partial
\delta
\phi_2 /
\partial t
\approx
\partial \delta \phi_3 / \partial t \approx 0$, one can express the fast modes
$\delta
\phi_2$ and $\delta \phi_3$ in terms of the slow mode, $\delta \phi_1$, and its
complex-conjugate, ${\delta
\phi _1}^{*}$. If now one  inserts the so-obtained expressions of the fast
modes
into the evolution equation of the slow mode, one obtains a closed nonlinear
equation for the latter ({\it adiabatic elimination} \cite{haken,nicolis}).  In
practice, however, such a calculation is possible only close to the the
bifurcation point, where the amplitude of
$\delta
\phi_1$ is supposed to approach zero as $R
\rightarrow R_c$. In fact, there exist other types of transitions, such as the
one arising in the Vanderpol equation, where the amplitude of the solution
above
the instability does not vanish as one approaches to the critical point
\cite{florence}.  Detailed analysis  shows that this is not the case here (i.e.
$|\delta
\phi_1| \rightarrow 0$ as $R
\rightarrow R_c$), so that we can limit ourselves to lowest orders in
$|\delta \phi_1|$, obtaining finally the so-called {\it normal form} or {\it
amplitude equation}
 for the slow mode:
\begin{eqnarray}
\frac{\partial \delta\phi_1(t)}{\partial t} & = &
\lambda \,\delta\phi_1(t)\nonumber\\ & - &
\gamma \,|\delta\phi_1(t)|^2\,\delta\phi_1(t)\,\left( 1
\,+\,O(|\delta\phi_1(t)|^2)
\right)\,,
\label{normalform}
\end{eqnarray}
where
\begin{equation}
\lambda \, \equiv \, \lambda_1(k_x = 1) \, = \,
 \frac{4 \pi^2}{R} \, \frac{a_r^2+1}{a_r^2 (a_r^2+2)} \,
\left(1 \, - \, \frac{R_c^2}{R^2}\right)
\,+
\,O\left( |R/R_c\,-\, 1|^2\right)\,,
\label{lambda}
\end{equation}
and $\gamma$ is a positive constant whose expression, to dominant
order in $|R/R_c\,-\, 1|$, is given by
\begin{equation}
\gamma  =  8\,\sqrt{2}\,\pi ^3\,
\frac{(a_r^6\,+\,17\,a_r^4\,+\,16\,a_r^2\,-\,32)\,(a_r^2\,+\,1)^2} {a_r^3
\,(a_r^2\,-\,1)^{3/2}\,(a_r^2\,+\,2)^3\,(a_r^2\,+\,4)^2}\,.
\label{gamma}
\end{equation}

Above the bifurcation point $R > R_c$ ($\lambda > 0$), the amplitude equation
(\ref{normalform}) admits two stable stationary solutions, corresponding to the
rotation sense of the stream lines in the fluid:
\begin{equation}
\delta\phi_1^{\pm} \, = \, \pm \, \sqrt{\frac{\lambda}{\gamma}} \,
\exp (i \,
\theta_0)\,,
\label{phistat}
\end{equation} where $\theta_0$ is a constant whose value depends on the
initial
conditions.  The fact that the stationary solution still depends on the initial
conditions simply reflects the Galilean invariance in the $x$ direction which
results from the periodic boundary conditions imposed on the system.  Using
relation (\ref{phistat}), one can compute the explicit form of the fast modes
for $k_{x}=\,0, \pm 1,\,\pm 2$. Applying the inverse transform
${\bf T}^{-1}(k_{x})$, cfr. eq. (\ref{Tmatrice}), to the so-obtained vector
${\bf
\delta\bphi}^{\pm}(k_{x})\,=\,
\left( \delta\phi_1^{\pm}, \delta\phi_2^{\pm}, \delta\phi_3^{\pm} \right)$ and
taking its inverse Fourier transform, one obtains the explicit expression
of the
stream function in real space. Up to order  $O(R/R_c\,-\, 1)$, one gets:
\begin{eqnarray}
\label{streamtheory}
\psi_{st}^{\pm} (x,\,y)& = & - \frac{1}{2 \,\pi} \cos(2\,\pi\,y)
 \, \pm \, \frac{R_c\,a_r}{2\,\pi \, (a_r^2 + 2 )}\, |\delta \phi_1| \, \bigg[
\cos (2 \pi x/a_r - \theta _0) \nonumber\\
 & &  - \, \frac{4\,\pi}{a_r \, R_c}\sin \left(2 \pi x/a_r -
 \theta_0 \right) \sin(2 \pi y) \bigg] \nonumber\\
 &  & + \,\frac{R_c^2}{2\,\pi \, (a_r^2 + 2 )^2} \,|\delta \phi_1|^2\,
 \bigg[1 \,- \,\frac{a_r^4}{(a_r^2+4)^2} \cos \left(4 \pi x/a_r - 2
\theta
 _0 \right) \bigg] \cos (2 \pi y)  \nonumber\\
\end{eqnarray} where we have set $|\delta \phi_1| \equiv |\delta
\phi_1^{\pm}|$.  Using  relations (\ref{streamfunction}), the velocity profiles
can now be obtained straightforwardly:
\begin{eqnarray}
u_{st}^{\pm}(x, \, y) & = & \sin(2 \pi y)
\, \mp  \,\frac{4 \,\pi }{(a_r^2+2)}\,|\delta \phi_1|\,
\sin \left(2 \pi x/a_r-\theta_0 \right) \cos(2 \pi y) \nonumber\\ & &
-\,\frac{R_c^2}{(a_r^2 + 2 )^2} \,|\delta \phi_1|^2\,
\bigg[1 \,- \,\frac{a_r^4}{(a_r^2 + 4)^2} \cos \left(4 \pi x/a_r - 2
\theta_0 \right) \bigg] \sin(2 \pi y) \,.
\label{ustationar}
\end{eqnarray}

\begin{eqnarray}
v_{st}^{\pm}(x,\,y) & = & \pm \, \frac{R_c}{(a_r^2 + 2 )}\,
|\delta \phi_1| \, \bigg[ \sin \left(2 \pi x/a_r - \theta  _0 \right) + \,
\frac{4\,\pi}{a_r \, R_c}\cos \left(2 \pi x/a_r -
 \theta_0 \right) \sin(2 \pi y) \bigg] \nonumber\\ & & -\, \frac{2\, R_c^2 \,
a_r^3}{(a_r^2 + 2 )^2\,(a_r^2+4)^2}
\,|\delta \phi_1|^2\,\sin
\left(4 \pi x/a_r - 2 \theta_0 \right) \cos(2 \pi y)\,.
\label{vstationar}
\end{eqnarray}

A density plot of the stream function (\ref{streamtheory}) is represented in
Figure (1) for $R = 15$, $a_r = 2$ and $\theta_0 = 0$ where, for the sake of
clarity, a vector plot of the velocity field is also included.  We note
that the
flow has an ABC-like topology \cite{usikov}, with closed streamlines (eddies),
open ones and separatrices  between them.

We recall that the above results rest on the "three modes" approximation
theory.
To check the validity of this basic assumption, we have solved numerically the
incompressible nonlinear hydrodynamic equations for $a_r = 2$, using standard
techniques \cite{chow}. Figure (2) compares contour
plots of the stream function obtained numerically with its corresponding
theoretical counterpart, eq. (\ref{streamtheory}), for
$R = 15$.  Given the relatively large distance from the critical point
($R/R_c\,-\, 1 \approx 17 \%$), the agreement is much better than expected, the
discrepancy remaining below $5 \%$. Surprisingly, the agreement does not
improve
as we consider smaller values of the Reynolds number.  This is shown in Figure
(3), where both the numerical and the theoretical horizontal profiles of the
stream function with a fixed value of the vertical coordinate,
$y=3/4$, are depicted for the Reynolds number  $R=13$.  The discrepancy now
exceeds $10 \%$.

To understand the origin of this unexpected behavior, we note that the value of
the critical Reynolds number that we have used to evaluate the stream function
(eq. \ref{streamtheory}) is based on the three modes approximation theory (cfr.
eq.
\ref{Rc}).  As shown before, the accuracy of the latter value of $R_c$ is about
$0.4 \%$, which is fine as far as the distance from the critical point
$(R/R_c\,-\, 1)$ remains much larger than  $0.4 \%$.  Now, for $R=13$, the
distance from the critical point is about $1 \%$ which is of the same order as
the accuracy of $R_c$ and explains the relatively important discrepancy we have
observed in Figure (3).

To overcome this difficulty, one has to compute a more accurate value of the
critical Reynolds number, based for instance on the 5 modes approximation
theory
(cfr. eq.
\ref{Rc5}).  As well known \cite{nicolis}, this correction concerns only the
value of $R_c$, and in no way compromises the validity of the amplitude
equation
(\ref{normalform}) and its corresponding solution eq.(\ref{streamtheory}).
This
is illustrated in Figure (3), where excellent agreement with the numerical
result is demonstrated, whenever we use
$R_c^{(5)}$ as the critical Reynolds number.  For smaller value of R, one
can as
well compute numerically the value of
$R_c$ with desired precision and used it as an input to the amplitude equation
(\ref{normalform}).

So far, we have limited ourselves to the analysis of the deterministic
equations
only, i.e. we have discarded the noise terms.  In principle, there is no
difficulty in taking into account the noise contributions as well, except that
the amplitudes of the field variables $\left( \delta\phi_1, \delta\phi_2,
\delta\phi_3 \right)$ are now directly related to the amplitude ${\cal B}$
of the noise, which is typically a small parameter. For example, the fast
variables
$\left(\delta\phi_2, \delta\phi_3 \right) \approx O({\cal B}^{\,1/2})$, whereas
the slow variable
$\delta\phi_1 \approx O({\cal B}^{\,1/4})$ (a detailed discussion of this
problem
is given in  \cite{chris}). Keeping this restriction in mind, one can
repeat all
the above calculations in the presence of noise terms. To the dominant order in
$|\delta \phi_1|$, one finds
\begin{eqnarray}
\frac{\partial \delta\phi_1(t)}{\partial t} & = &
\lambda \,\delta\phi_1(t) \, - \,
\gamma \,|\delta\phi_1(t)|^2\,\delta\phi_1(t)\,+\, \xi(t)\,,\nonumber \\
\frac{\partial \delta\phi_1^*(t)}{\partial t} & = &
\lambda \,\delta\phi_1^*(t) \, - \,
\gamma \,|\delta\phi_1(t)|^2\,\delta\phi_1^*(t)\,+\, \xi^*(t)\,.
\label{normalformnoise}
\end{eqnarray}
The functions
$\xi(t)$ and its complex-conjugate $\xi^*(t)$ are Gaussian white noises with
zero means and correlations given by:
\begin{eqnarray} <\xi(t)\,\xi(t')> & = & 0 \nonumber\\
<\xi(t)\, \xi ^{*}
(t')> & = & {\cal B}\,\delta(t\,-\,t')
\label{covarxi}
\end{eqnarray}
with
\begin{eqnarray}
{\cal B} \, = \, \frac{4\,k_{B}\,T_0\,a_r^2}
{M \, u_0^2 \, R} \bigg[1\,+\,
O\left(|R/R_{c}-1| \right) \bigg]\,,
\label{valB}
\end{eqnarray}
$M$ being the total mass of the system:
\begin{equation}
\label{masse}
M \, = \, a_r \rho_0 L_y^2 \,.
\end{equation}

The results derived in this section were based explicitly on the
incompressibility assumption.  However, as discussed in the Introduction,  this
assumption is inconsistent with the very foundation of the
fluctuating hydrodynamic formalism.  On the other hand, we have presented in
\cite{malek1} numerical evidence that in the vicinity of the bifurcation point
the system behaves basically as an incompressible fluid.  We therefore expect
that the Langevin equation (\ref{normalformnoise}) should remain valid for $R$
close enough to $R_c$. We shall clarify this main issue in the next section.\\

\section{Fluctuations in the compressible flow}
\label{sec:III}

Let us now consider the compressible hydrodynamic equations (\ref{continuity} -
\ref{stress}) for which we need to specify an equation of state.  Since the
system is isothermal, we simply set
\begin{equation} p\,=\,c_s^2\,\rho\,,
\end{equation} where $c_s$ is the isothermal speed of sound.  As in the
previous
section, we start with the linearized hydrodynamic equations around the
reference state \{$\rho_0, {\bf v}_{st}$\}, where ${\bf v}_{st}$ is given by
eq.(\ref{vstat}).  Setting
\begin{eqnarray}
\label{drodv}
\rho & = & \rho_0\,+\,\delta\rho\,,\nonumber\\ {\bf v} & = & {\bf
v}_{st}\,+\,\delta{\bf v}\,,
\end{eqnarray} and scaling lengths by
$L_y$, time by
$L_y/c_s$, $\delta \rho$ by $\rho_0$ and $\delta {\bf   v}$ by the speed of
sound $c_s$, the dimensionless linear fluctuating equations in the Fourier
space
read (recall that $\tilde{k}_x \equiv k_x/a_r$):
\begin{eqnarray}
\label{rofour}
\frac{\partial\,\delta\rho_{k_x,\,k_y}(t)}{\partial t} & = & 2\,\pi\,i\,
\left( \tilde{k}_x\,\delta u_{k_x,\,k_y}\,+\,\,k_y\,\delta v_{k_x,\,k_y}
\right)
\, + \,
\varepsilon\,R\,\pi\,\tilde{k}_x\,(\delta\rho_{k_x,\,k_y+1}\,-\,\delta
\rho_{k_x,\,k _y-1})\,,
\end{eqnarray}
\begin{eqnarray}
\frac{\partial \delta u_{k_x,\,k_y}(t)}{\partial t} &  =  & -
\pi\,\varepsilon\,R(\delta v_{k_x,\,k_y+1}\,+\,\delta v_{k_x,\,k_y-1}) \, + \,
\pi\,\varepsilon\,R\,\tilde {k}_x\,(\delta u_{k_x,\,k_y+1}\,-\,\delta
u_{k_x,\,k_y-1})
\nonumber\\
   & - &  \,4\,\pi^2\,\varepsilon({\tilde {k}_x}^2\,+\,k_y^2)\,\delta
u_{k_x,\,k_y} \, -
\,4\,\pi^2\,\alpha\,\varepsilon\,\tilde {k}_x (\tilde {k}_x\,\delta
u_{k_x,\,k_y}\,+\, k_y\,\delta v_{k_x,\,k_y}) \nonumber\\
   & + & 2\,\pi\,i\,\tilde{k}_x\,\delta\rho_{k_x,\,k_y}\, + F_{k_x,\,k_y}(t)\,,
\end{eqnarray}
\begin{eqnarray}
\frac{\partial \delta v_{k_x,\,k_y}(t)}{\partial t}  & = &
\pi\,\varepsilon\,R\,\tilde {k}_x\,(\delta v_{k_x,\,k_y+1}\,-\,\delta
v_{k_x,\,k_y-1})
\,-\,4\,\pi^2\,\varepsilon({\tilde {k}_x}^2\,+\,k_y^2)\,\delta v_{k_x,\,k_y}
\nonumber \\
 & - & 4\,\pi^2\,\alpha\,\varepsilon\, k_y\,(\tilde {k}_x\, \delta
u_{k_x,\,k_y}\,+\,k_y\,\delta v_{k_x,\,k_y})\, +
2\,\pi\,i\,k_y\,\delta\rho_{k_x,\,k_y} \, + \, G_{k_x,\,k_y}(t)\,,
\label{complin}
\end{eqnarray}
 where $R$ is the Reynolds number, defined in eq.(\ref{defR}),
\begin{equation}
\varepsilon \,=\, \frac{\eta}{\rho _0\,c_s\,L_y} \,,
\label{epsilon}
\end{equation} and
\begin{equation}
\alpha \,=\, \zeta / \eta\, .
\end{equation}
The functions $F_{k_x,\,k_y}$ and $G_{k_x,\,k_y}$ are Fourier components of the
noise terms ; their covariances follow directly from eqs. (\ref{stress},
\ref{noisecor}):
\begin{eqnarray}
< F_{k_x,\,k_y}(t)\,F_{k'_x,\,k'_y}(t') > & = &  8\,\pi ^2 \, \varepsilon \,
{\cal A}\,[(\alpha\,+\,1)\,
\tilde{k}_x^2\,+\,k_y^2]\, \delta_{{\bf   k}+{\bf k}',0}^{Kr}\,\delta
(t\,-\,t')\,,\nonumber\\
< F_{k_x,\,k_y}(t)\,G_{k'_x,\,k'_y}(t') > & = &  8\,\pi ^2 \,
\varepsilon \, {\cal A}\,\alpha\,
\tilde{k}_x\,k_y \, \delta_{{\bf   k}+{\bf   k}',0}^{Kr}\,\delta (t\,-\,t')\,,
\nonumber\\
<G_{k_x,\,k_y}(t)\,G_{k'_x,\,k'_y}(t') > & = &  8\,\pi ^2 \, \varepsilon \,
{\cal
A}\,[{\tilde {k}_x}^2\,+\,(\alpha\,+\,1)\,k_y^2]\, \delta_{{\bf   k}+{\bf
k}',0}^{Kr}\,\delta (t\,-\,t')\, , \nonumber\\
\label{fourcovar}
\end{eqnarray}
where ${\bf   k} \equiv (\tilde{k}_x , \, k_y)$ and
\begin{equation}
{\cal A}\,=\, \frac{k_B\,T_0}{M \,{c_s}^2} \, ,
\label{valA}
\end{equation}
$M$ being the total mass of the system (cfr. eq. (\ref{masse})).

For the sake of clarity, we first focus on the {\it deterministic} behavior,
i.e. we discard for the moment the noise contributions from the evolution
equations (\ref{rofour} -
\ref{complin}).  Furthermore, we shall limit ourselves to the 3-modes
approximation theory, i.e. we shall neglect the modes with
$|k_y|\, \ge \,2 $, for the very same reasons that we have discussed for the
incompressible case.  With these assumptions,  eqs. (\ref{rofour} -
\ref{complin}) reduce to a system of nine coupled equations.  It can then be
checked that the change of variables
\begin{eqnarray}
\label{changevar}
\delta\rho_{k_x}^{\pm}(t) & = &
\delta\rho_{k_x,\,1}(t)\, \pm \,\delta\rho_{k_x,\,-1}(t)\nonumber\\
\delta u_{k_x}^{\pm}(t) & = & \delta u_{k_x,\,1}(t)\, \pm \,\delta u
_{k_x,\,-1}(t)\,\nonumber\\
\delta v_{k_x}^{\pm}(t) & = & \delta v_{k_x,\,1}(t)\,\pm \,\delta
v_{k_x,\,-1}(t)
\end{eqnarray} leads to a "partial diagonalization" of the evolution equations,
i.e. the equations for the variables
\{$\delta\rho_{k_x, 0}$, $\delta\rho_{k_x}^{-}$, $\delta u_{k_x, 0}$, $\delta
u_{k_x}^{-}$, $\delta v_{k_x}^{+}$\} decouple from the rest.  Furthermore,
their
associated eigenvalues prove to remain strictly negative, regardless of the
value of the Reynolds number $R$, so that they are not determinant for the
onset
of convective instability.  We therefore focus on the remaining four variables
$\left\{\delta\rho_{k_x}^{+}
\delta u_{k_x}^{+} , \delta v_{k_x}^{-} , \delta v_{k_x,0} \right\}$. Defining
the vector $\delta {\bf h}_{k_x}(t) \equiv \left\{\delta\rho_{k_x}^{+},\,
\delta u_{k_x}^{+} ,\, \delta v_{k_x}^{-} ,\, \delta v_{k_x,0} \right\}$ one
readily finds:
\begin{equation}
\frac{\partial}{\partial t}
\delta {\bf h}_{k_x}(t)\,=\,{\bf C}(k_x)\,\cdot
\delta {\bf h}_{k_x}(t)\,,
\end{equation} where the matrix ${\bf C}$ is given by:
\begin{eqnarray} &  & {\bf C}(k_x)\,=\,\nonumber\\ &  & \left(
\begin{array}{cccc} 0 & 2 \pi i \tilde{k}_x & 2 \pi i & 0 \\ 2 \pi i
\tilde{k}_x
& - 4 \pi^2 \varepsilon (1+ \alpha \tilde{k}_x^2 +
\tilde{k}_x^2)
 & -4 \pi^2 \varepsilon \alpha \tilde{k}_x  & -2 \pi \varepsilon R\\
 2 \pi i & -4 \pi^2 \varepsilon \alpha \tilde{k}_x
 & -4 \pi^2 \varepsilon (1 + \alpha + \tilde{k}_x^2) & -2 \pi \varepsilon
\tilde{k}_x R\\
 0 & 0 & \pi \varepsilon \tilde{k}_x R & -4 \pi^2 \varepsilon \tilde{k}_x^2
\end{array}
\right)\,.\nonumber\\
\end{eqnarray}

The analysis can be simplified somewhat by noticing that the  parameter
$\varepsilon$ must remain  small if one wishes to remain within the limit of
validity of the hydrodynamic regime
\cite{alley}. Furthermore, as already mentioned in the introduction, in this
article we limit ourselves to strictly sub-sonic flows, so that we shall
restrict the analysis to a parameter domain where
\begin{equation}
\label{Repsilon}
\varepsilon \, \ll \,1 \,\,\,\,\,\,\,\, ; \,\,\, \,\,\,\, \varepsilon R \,=\,
u_0/c_s
\, \ll
\,1.
\end{equation}
Accordingly, we evaluate the eigenvalues of the matrix ${\bf C}$
perturbatively:
\begin{equation}
\tilde{\lambda}(k_x)\,=\,\tilde{\lambda}^{(0)}(k_x)\,+\,\varepsilon\,
\tilde{\lambda}^{(1)}(k_x)\,+
\, \dots
\end{equation} After some algebra, one finds, up to order $O(\varepsilon ^2)$:
\begin{eqnarray}
\tilde{\lambda}_1(k_x) & = & \varepsilon
\left[-2\,\pi^2(1\,+\,2
{\tilde{k}_x}^2)\,+\,\pi\,\sqrt{4\,\pi^2\,+\,2\,R^2\,\tilde{k_x}^2\,(1\,-\,{
\tilde{k}_x}^2)/(1\,+\,{\tilde {k}_x}^2)} \,\,\, \right]\,,\nonumber\\
\tilde{\lambda}_2(k_x) & = & \varepsilon
\left[-2\,\pi^2(1\,+\,2{\tilde{k}_x}^2)\,-\,\pi\,
\sqrt{4\,\pi^2\,+\,2\,R^2\,\tilde{k_x}^2\,(1\,-\,{\tilde{k}_x}^2)/(1\,+\,{\tilde
{k}_x}^2)}
 \,\,\, \right]\,,\nonumber\\
\tilde{\lambda}_3(k_x) & = & 2\,\pi\,i\,\sqrt{1\,+\,{\tilde
{k}_x}^2}\,-\,2\,\pi^2(\alpha\,+1)\,\varepsilon(1\,+\,{\tilde
{k}_x}^2)\,,\nonumber\\
\tilde{\lambda}_4(k_x) & = & -2\,\pi\,i\,\sqrt{1\,+\,{\tilde
{k}_x}^2}\,-\,2\,\pi^2(\alpha\,+1)\,\varepsilon(1\,+\,{\tilde {k}_x}^2)\,.
\label{eigenval}
\end{eqnarray}
The eigenvalues $\tilde{\lambda}_1$ and
$\tilde{\lambda}_2$ correspond to dissipative (viscous) modes, while
$\tilde{\lambda}_3$ and $\tilde{\lambda}_4$  are related to the propagation of
(damped) sound waves. It can then be easily checked that the real 
parts of
$\tilde{\lambda}_2,\,\tilde{\lambda}_3$ and $\tilde{\lambda}_4$  are always
negative, whereas there exists a critical value of the Reynolds number
\begin{equation}
\label{reycritcomp} R_c(k_x)\,=\,2\,\sqrt{2}\,\pi\,\frac{1\,+\,{\tilde
{k}_x}^2}{\sqrt{1\,-\,{\tilde {k}_x}^2}}\,
\,\,\,\,\,\,\,;\,\,\,\,\,\,\,0\,<\, {\tilde{k}_x}^2\,<1\,,
\end{equation} for which $\tilde{\lambda}_1$ vanishes, thus indicating the
limit
of stability of the corresponding mode.

Remarkably, the above expression of the critical Reynolds number is
identical to
the one obtained in the incompressible case (cfr. eq.(\ref{Rc})).  In fact,
detailed analysis shows that the relation (\ref{reycritcomp}) is exact,
i.e. it is
independent of
$\varepsilon$, at least within the framework of the 3-modes approximation
theory. On the other hand, if the modes $k_y = \pm 2$ are taken into account as
well, tedious calculations lead to
\begin{equation} R_c^{(5)}(k_x)\,=\,R_c(k_x)\,\left[1 \, + \frac{{\tilde
{k}_x}^4 \, ({\tilde {k}_x}^2 + 3)}{2\, ({\tilde {k}_x}^2+4)^{2}\,({\tilde
{k}_x}^2-1)}\right]^{- 1/2} +\,{\cal O}\left( (u_0/c_s)^2 \right)
\,  ; \, \,  \, 0 \, < \, {\tilde {k}_x}^2 \, < \, 1
\label{Rc5c}
\end{equation} which is again equivalent to the corresponding result obtained
for the incompressible case, eq. (\ref{Rc5}), the correction being of the order
of $O
\big(\varepsilon^2\big)$. In particular, the first mode to become unstable
corresponds to $|k_x| = 1$, provided $a_r > 1$.

We note that the matrix ${\bf C}$ is singular for
$k_x\,=\,0$, i.e. one of its eigenvalues vanishes.  A close inspection shows
that this zero eigenvalue corresponds to the mode
$\delta v_{0,0}$ which is identically zero because of linear momentum
conservation.  Accordingly, in what follows we shall concentrate on the case
$k_x\,\neq \,0$, looking for a similarity transformation ${\bf S}
\cdot {\bf C} \cdot {\bf S}^{-1}$ which diagonalizes the matrix ${\bf C}$. For
consistency, here again we perform the calculations perturbatively, i.e. we
expand
${\bf S}$ in powers of $\varepsilon$:
\begin{equation}
 {\bf S}(k_x)\,=\,{\bf S_0}(k_x)\,+\,\varepsilon\,{\bf S_1}(k_x)\, + \,
\dots
\end{equation} Note that this method constitutes an alternative to the time
scale perturbation theory
\cite{titulaer} that was generalized by Schmitz and Cohen \cite{cohen} in order
to study the B\'{e}nard instability in  a compressible fluid.

Since the explicit form of the eigenvalues are known up to
$O(\varepsilon^2)$, we only need to evaluate ${\bf S}$ (and its inverse ${\bf
S}^{-1}$) up to the same order.   Despite this simplification, the  general
expression of ${\bf S}$ is quite awkward and will not be presented here.  The
rest of the calculations are quite straightforward, but remain tedious and
lengthy, so we only give a brief sketch of the basic steps (see the discussions
below eq. (\ref{phi1}) ).

We start by taking the Fourier transform of the fluctuating hydrodynamic
equations (\ref{continuity}~-~\ref{stress}).  Using the change of variables
(\ref{drodv}) and (\ref{changevar}), we next derive the nonlinear fluctuating
equations for
$\delta {\bf h}_{k_x}$.  We then apply the transformation ${\bf S}$ to the
latter, obtaining a set of four nonlinear equations for the variables
$\left\{\delta \tilde{\phi}_1,
\delta \tilde{\phi_2} , \delta \tilde{\phi_3} , \delta \tilde{\phi_4}
\right\} \equiv \delta\tilde{{\bphi}}(t) = {\bf S} \cdot \delta {\bf
h}_{k_x}$.   Close to the bifurcation point ($R \approx R_c$ ,
$k_x = 1$), the mode $\delta \tilde{\phi}_1$ exhibits a critical slowing down,
since by construction $\tilde{\lambda}_1 \approx 0$.  We can therefore
proceed to
an adiabatic eliminations of the "fast" modes $\left\{\delta \tilde{\phi_2} ,
\delta \tilde{\phi_3} , \delta \tilde{\phi_4}
\right\}$, limiting ourselves to dominant orders in
$|\delta
\tilde{\phi}_1|$ (see the paragraph preceding eq. (\ref{normalformnoise})
). The
final result is a set of two coupled Langevin equations for the slow mode
$\delta
\tilde{\phi}_1$ and its complex conjugate $\delta
\tilde{\phi}_1^*$ :
\begin{eqnarray}
\frac{\partial \delta \tilde{\phi}_1(t)}{\partial t} & = &
\tilde{\lambda} \,\delta \tilde{\phi}_1(t)\, - \,
\tilde{\gamma} \,|\delta \tilde{\phi}_1(t)|^2 \,\delta \tilde{\phi}_1(t)\,+\,
\tilde{\xi}(t) \nonumber\\
\frac{\partial \delta \tilde{\phi}_1^*(t)}{\partial t} & = &
\,\tilde{\lambda} \delta \tilde{\phi}_1^*(t)\,-
\,\tilde{\gamma} \, |\delta \tilde{\phi}_1(t)|^2  \, \delta
\tilde{\phi}_1^*(t)\,+\tilde{\xi}^{*}(t)
\label{nfsmcc}
\end{eqnarray}
with
\begin{equation}
\tilde{\lambda} \,=\, \tilde{\lambda}_1(k_x=1)\,=
\,\lambda \,\, \frac{u_0}{c_s} \,\left[1\,+\,{\cal O}(u_0^2/c_s^2)\,\right] \,
\approx  \, 4 \pi^2 \varepsilon \, \frac{a_r^2+1}{a_r^2 (a_r^2+2)} \,
\left(1 \, - \, \frac{R_c^2}{R^2}\right)
\label{tildelambda}
\end{equation}
and
\begin{eqnarray}
\tilde{\gamma} =  \gamma\,\,\frac{c_s}{u_0}\,
\left[1\,+\,{\cal O}(u_0^2/c_s^2)\,\right] \, \approx \,
\frac{\gamma}{\varepsilon \, R}
\label{tildegamma}
\end{eqnarray}
where $\lambda$ and $\gamma$ are given by eqs. (\ref{lambda}) and
(\ref{gamma}), respectively.  The functions
$\tilde{\xi}(t)$ and its complex-conjugate $\tilde{\xi}^*(t)$ are Gaussian
white
noises with zero means and correlations given by:
\begin{eqnarray} <\tilde{\xi}(t)\,\tilde{\xi}(t')> & = & 0\,,\nonumber\\
<\tilde{\xi}(t)\, \tilde{\xi}^*(t')> & = & \tilde{\cal B}\,\delta(t\,-\,t')\,,
\label{compcovarxi}
\end{eqnarray}
 with
\begin{eqnarray}
\tilde{\cal B}  =  \left(\frac{u_0}{c_s}\right)^3  {\cal B}\,
\left[1\,+\,{\cal O}(u_0^2/c_s^2)\,\right] \, \approx \, 4 \, \varepsilon
\, a_{r}^2 \,
{\cal A} \, ,
\label{compvalA}
\end{eqnarray}
where ${\cal B}$ and ${\cal A}$ are given by eqs. (\ref{valB}) and
(\ref{valA}), respectively.

Although the form of the Langevin equations (\ref{nfsmcc}) is the same as
the one
obtained for the incompressible case, eqs. (\ref{normalformnoise}), they are
nevertheless not equivalent since their coefficients are clearly different,
even
to dominant order in $\varepsilon$.    The main reason for this apparent
discrepancy is related to the fact that, for the incompressible case, the
analysis has been carried out by scaling the velocities by $u_0$, whereas for
the compressible case we have used a different scaling, i.e. we have scaled the
velocities by the velocity of sound
$c_s$. If now we switch back to the former scaling, i.e. we perform the change
of variables $ t
\,
\rightarrow \, t \,\, c_s / u_0$, $\{u, v\} \, \rightarrow \,  u_0 / c_s \{u,
v\}$, then eqs. (\ref{nfsmcc}) lead to
\begin{eqnarray}
\delta \tilde{\phi}_1(t) \, = \, \frac{u_0}{c_s} \,\, \delta \phi_1(t) \,
\left[1\,+\,{\cal O}(u_0^2/c_s^2)\,\right]
\label{phiscale}
\end{eqnarray}
Remarkably, this result shows that, to dominant order in $\varepsilon$, the
evolution of fluctuating compressible and incompressible hydrodynamic equations
are governed by the very same slow mode, at least for
values of Reynolds number close to its critical value.

Let us first consider the macroscopic behavior.  Using eqs.
(\ref{tildelambda}), (\ref{tildegamma}) and (\ref{phiscale}), one can go
backward step by step
and derive as well the evolution equations of the hydrodynamical velocities
near the
instability threshold.  It can then be easily checked that, to dominant
order in
$\varepsilon$, the compressible stationary velocity profiles are given by their
incompressible expressions, eqs. (\ref{ustationar} and \ref{vstationar}). To
check this important result, we have solved numerically the full nonlinear
compressible hydrodynamic equations and compared the result with analytical
expressions obtained for the incompressible case.  A typical result is shown in
Figure (4), where
$u_{st}(x, y=1/4)$ as a function of $v_{st}(x, y=1/4)$ is depicted for $R=15$,
 $\varepsilon = 10^{-2}$ and $a_r = 2$. Given the relatively large values of
the Reynolds number ($R / R_c - 1 \approx 17 \% $) and $\varepsilon$, the
agreement is very good, the discrepancy remaining below  $5 \%$.

We now concentrate on the behavior of fluctuations, as described by the
Langevin
equations (\ref{nfsmcc}).  The associated Fokker-Planck equation reads:
\begin{eqnarray}
\frac{\partial P(\delta {\tilde{\phi}}_1\,,\delta
{\tilde{\phi}}_1^{*}\,,t)}{\partial t}
\, = & &   \frac{\partial}{\partial (\delta {\tilde{\phi}}_1)}
\left[-({\tilde{\lambda}}
\,\delta {\tilde{\phi}}_1\,-\,{\tilde{ \gamma}} \delta {\tilde{\phi}}_1
^2\,\delta
{\tilde{\phi}}_1^{*})\,P\,+\,\frac{\tilde{{\cal B}}}{2} \frac{\partial
P}{\partial
(\delta {\tilde{\phi}}_1^{*})} \right] \nonumber\\ & + &
\frac{\partial}{\partial (\delta {\tilde{\phi}}_1^{*})}
\left[-({\tilde{\lambda}}
\,\delta {\tilde{\phi}}_1^{*}\,-\,{\tilde{ \gamma}} \delta
{\tilde{\phi}}_1\,{\delta
{\tilde{\phi}}_1^{*}}^{2})\,P\,+\,\frac{\tilde{{\cal B}}}{2} \frac{\partial
P}{\partial (\delta {\tilde{\phi}}_1)} \right] \,.
\end{eqnarray} At the stationary state, one finds:
\begin{equation}
\label{pstation}
P_{st}(\delta {\tilde{\phi}}_1\,,\delta
{\tilde{\phi}}_1^{*}) = {\cal N}^{-1} \, \exp \left[\frac{2}{\tilde{{\cal
B}}} \left(
{\tilde{\lambda}}|\delta
{\tilde{\phi}}_1|^2\,-\frac{{\tilde{ \gamma}}}{2} |\delta
{\tilde{\phi}}_1|^4 \right)
\right]
\end{equation}
with
\begin{equation}
{\cal N} \, = \, \frac{1}{4} \,\sqrt{\pi \, \tilde{{\cal
B}}/\tilde{\gamma}\,\,}
\,
\exp\left(\tilde{\lambda}^2/\,\tilde{\gamma}\tilde{{\cal B}} \right) \,\, {\rm
erfc}\left(- \, \tilde{\lambda} \,/\sqrt{\tilde{\gamma}\,\tilde{{\cal
B}}\,}\,\right)
\end{equation}
where ${\rm erfc}(\dots)$ stands for the complementary error function.
Thanks to this
result, one readily gets:
\begin{equation}
\label{delphiNL}
<|\delta {\tilde{\phi}}_1|^2> \, = \, \frac{1}{\tilde{\gamma}} \,
\left(\tilde{\lambda} \, + \, \tilde{{\cal B}}\, / 4 \,{\cal N}\, \right)
\end{equation}

Away from the
bifurcation point (${\tilde{\lambda}} << 0$) the quartic term in
(\ref{pstation}) is
negligible so that the distribution is Gaussian and
\begin{equation}
<|\delta {\tilde{\phi}}_1|^2>_{_G} \,\, \approx \,\frac {{\cal A} \, R^2 \,
a_r^4 \,
(a_r^2+2)}{2 \, \pi^2
\left(R_ c^{2}-R^2 \right ) \left (a_r^2 \, + \, 1
\right )}
\label{corphiG}
\end{equation}
The fluctuations thus behave as $|\delta
{\tilde{\phi}}_1| \, \approx \, {\cal O}({\cal A}^{1/2})$.  Recall that the
parameter
${\cal A}$ is inversely proportional to the system's total number of
particles so that ${\cal
A} \, << \, 1$ (cfr.
eq. (\ref{valA})). As one approaches the bifurcation point, the Gaussian
character of the
distribution is gradually lost.  Right at the bifurcation point,
${\tilde{\lambda}} = 0$, one has
\begin{equation} <|\delta {\tilde{\phi}}_1|^2>_{\tilde{\lambda}=0} \,\, = \, 2
\, \varepsilon \,
a_r
\,
\left(
\frac{ R_c \, {\cal A} }{ \gamma\, \pi} \right)^{1/2}
\end{equation}
which shows that the fluctuations now behave as $|\delta
{\tilde{\phi}}_1| \, \approx \, {\cal O}({\cal A}^{1/4})$. The
enhancement of fluctuations and the change of the probability law at the
bifurcation point are a direct manifestation of spatial symmetry breaking
associated with the emergence of convective patterns.

On the other hand, the fast modes $\left\{\delta \tilde{\phi_2} ,
\delta
\tilde{\phi_3} , \delta
\tilde{\phi_4}
\right\}$ prove to remain Gaussian, regardless of the value of the Reynolds
number.  Detailed analysis shows that their contribution to nonequilibrium
statistical
properties of the fluid remain of the order of ${\cal O}(u_0^2/c_s^2)$.  In
other words,
the fluctuation spectrum of hydrodynamic variables are mainly determined
by the statistical properties of $\delta \tilde{ \phi_1}$. For instance,
the static velocity
auto-correlation function is found to obey:
\begin{equation}
\label{delvNL}
<\delta {\bf   v}_{{\bf   k} }  \cdot  \delta {\bf   v}_{-{\bf   k} }> \, -
\, 2 {\cal A} \,  =  \, \frac{\pi ^{2}
(a_{r}^{2}+1)}{a_{r}^{2}\,(a_{r}^{2}+2)^{2}} \, <|\delta
{\tilde{\phi}}_1|^2> \, \bigg[1 \,
+\,{\cal O}\left( (u_0/c_s)^2 \right) \bigg]
\end{equation}
where the second term on the left hand side is the equilibrium contribution
and  $<|\delta
{\tilde{\phi}}_1|^2>$ is given by eq. (\ref{delphiNL}).

It is instructive to study the
Gaussian limit,  $R << R_c$, where the linearized Langevin equations, eqs.
(\ref{rofour} -
\ref{complin}), remain valid.  As has been shown in \cite{malek1}, they
lead to the
following expression for the static velocity
auto-correlation function:
\begin{equation}
<\delta {\bf   v}_{{\bf   k} }  \cdot  \delta {\bf   v}_{-{\bf   k}
}>_{_{\rm G}} \, -
\, 2 {\cal A} \,  =  \,\frac {{\cal A} \, R^2 \, a_r^2}{2
\left(R_ c^{2}-R^2 \right ) \left (a_r^2 \, + \, 2
\right )}
\label{delvLin}
\end{equation}
Now, inserting into eq. (\ref{delvNL}) the Gaussian form of $<|\delta
{\tilde{\phi}}_1|^2>$,
as given by eq. (\ref{corphiG}), leads precisely to the very same result.
We thus conclude
that our general expression, eq. (\ref{delvNL}), remains valid in the
Gaussian regime, $R
<< R_c$, despite the fact that it has been derived in the close vicinity of the
bifurcation point $R
\approx R_c$.

To check the validity of our theoretical results, we have simulated
the nonlinear fluctuating hydrodynamic equations (\ref{continuity} -
\ref{stress}) for different values of
$R$, setting $a_r = 2$, $\varepsilon = 10^{-2}$ and ${\cal A} = 10^{-3} /
256 \approx 3.9
\times 10^{-6}$.  The estimated statistical errors remain below $5\%$ for
$R \le 10$, but
grows rapidly as we consider higher values of $R$, reaching about $13 \%$
for $R \approx
R_c$. Above the bifurcation point, $R > R_c$, the stationary distribution
has two
maxima, located at
$\delta {\tilde{\phi}_1}\,=\, \pm \sqrt{{\tilde{\lambda}} / {\tilde{
\gamma}}}$, which
correspond (up to a phase factor) to the deterministic stationary solutions of
the amplitude equation (\ref{phistat}).  Because of the presence of noise
terms, the system
visits these states in a rather random fashion, resulting into a huge
dispersion of data.  This
is specially true for $R$ close to $R_c$, which is precisely the situation
where our
theoretical predictions are expected to be applicable. Under this
circumstance, obtaining
reliable statistics requires prohibitively large computing times, so that
we have been forced
to limit the numerical simulations to values of Reynolds numbers $R
\le R_c$.

The results are presented in Figure (5), together with both  the complete
and the
linearized solutions, eqs. (\ref{delvNL}) and (\ref{delvLin}),
respectively. The
linear theory (Gaussian limit) shows quantitative agreement for values of
$R / R_c$ up
to about $86 \%$, but significant discrepancies start to show up as $R
\rightarrow R_c$ where
the theory leads to diverging correlation functions (cfr. eq.
(\ref{delvLin})).  This is not
the case for the complete solution, eq. (\ref{delvNL}), which exhibits
perfect quantitative
agreement for $R / R_c$ up to
$95 \%$. A relatively small discrepancy of about $8 \%$ is however observed
for higher values
of
$R$.  Although this discrepancy remains within the limit of the estimated
statistical errors,
its systematic aspect requires nevertheless some clarifications.  In this
respect, it is
important to recall that  the results derived in this
section were valid up
to ${\cal O}(u_0^2/c_s^2)$.  Now, by definition $u_0/c_s = R \,
\varepsilon$ (cfr. eq.
(\ref{Repsilon})), and since we have set $\varepsilon = 10^{-2}$,  $ R_c \,
\varepsilon
\approx 0.13 $ at the bifurcation point.  This relatively large value of $
R_c \,
\varepsilon$ might well be
at the origin of the observed discrepancy.  To check the validity of this
argument, it is
tempting to perform the simulations all over again for a smaller value of
$\varepsilon$.
However, since the relaxation time of hydrodynamical modes grows as
$\varepsilon^{-1}$, reaching the same degree of statistical  accuracy as
for the previous
cases requires much more longer running times.  For this reason we
decided to perform only one more simulation right at the critical point,
$R=R_c$, setting
$\varepsilon = 10^{-3}$. The theoretical prediction for the nonequilibrium
part of the
velocity correlation function is $2.31 \times 10^{-6}$.  The simulation
leads to $2.24
\times 10^{-6}$ with an estimated statistical errors of about $15 \%$.  The
discrepancy
is now about
$3
\%$, much better than for the case
$\varepsilon = 10^{-2}$.

\newpage
\section{Concluding remarks}
\label{sec:IV}

Recently, we have studied the
statistical properties of the linearized Kolmogorov flow, from near
equilibrium up to the vicinity of the first instability leading to the
formation of vortices \cite{malek1}. In particular, we have established
that the
incompressibility assumption leads to  a wrong  form of the
static correlation functions, except near the instability threshold where
numerical results suggest that the incompressibility assumption should
remain valid. The clarification of this important issue requires a nonlinear
analysis of the fluctuating Kolmogorov flow. This is precisely the main purpose
of the present article.

We have first considered the case of an incompressible fluid.  After
identifying
the slow modes, governing the evolution of the system in the vicinity of the
instability threshold, we have performed  an adiabatic elimination of the
fast modes to obtain a set of two nonlinear Langevin equations for the slow
modes.  We have then succeeded to derive the explicit form of the
stationary stream
function, as well as the corresponding velocity profiles, in real space.
Numerical studies of the nonlinear hydrodynamical equations allowed us to
confirm
our theoretical predictions.

We have next considered the case of the compressible Kolmogorov flow.  The
analysis can be simplified somewhat by noticing that the evolution of
a compressible fluid is generally characterized by two different time
scales:  a
slow one, related to the dissipative viscous modes, and a fast one,
expressing the propagation of (damped) sound modes.  The ratio of these time
scales, denoted by $\varepsilon$ (cfr. eq. \ref{epsilon}), can be
considered as a
small parameter, since otherwise the very validity of the hydrodynamics can no
more be guaranteed \cite{alley}.  We thus have at our disposal a natural small
parameter which can be used to set up a perturbative technique. As already
mentioned, this method constitutes an alternative to the time
scale perturbation theory that was generalized by Schmitz and Cohen
in order to study the B\'{e}nard instability in a compressible fluid
\cite{cohen,titulaer}.

Using this perturbation technique, we have first shown that the
macroscopic behavior of the fluid is not affected - up to ${\cal
O}(u_0^2/c_s^2)$
- by the compressibility, in agreement with the intuitive arguments that we
have
presented in the Introduction.  We then succeeded to establish that, close to
the instability threshold, the stochastic dynamics of the system is governed by
two coupled non linear Langevin equations in the Fourier space.  The
solution of
these equations can be cast into the exponential of a Landau Ginzburg
functional which, to dominant order in $\varepsilon$, proves to be identical to
the one obtained for the case of the incompressible fluid.  The
theoretical predictions have been confirmed by numerical simulations of the
nonlinear fluctuating hydrodynamic equations.

\acknowledgments{We are very grateful to professors E. G. D. Cohen, G. Nicolis,
J. W. Turner and C. Van den Broeck for helpful comments. This work is supported
by the Belgian Federal Office for Scientific, Technical and Cultural Affairs
within the framework of the ``P\^ oles d'attractions interuniversitaires"
program, and by an European Commission DG 12 Grant PSS*1045.}

\begin{figure}
\label{fig1}
\caption{ Density plot of the stream function, eq. (\ref{streamtheory}),
for $R =
15$, $a_r = 2$ and
$\theta_0 = 0$. For the sake of clarity, a vector plot of the velocity
field is also included.}
\end{figure}

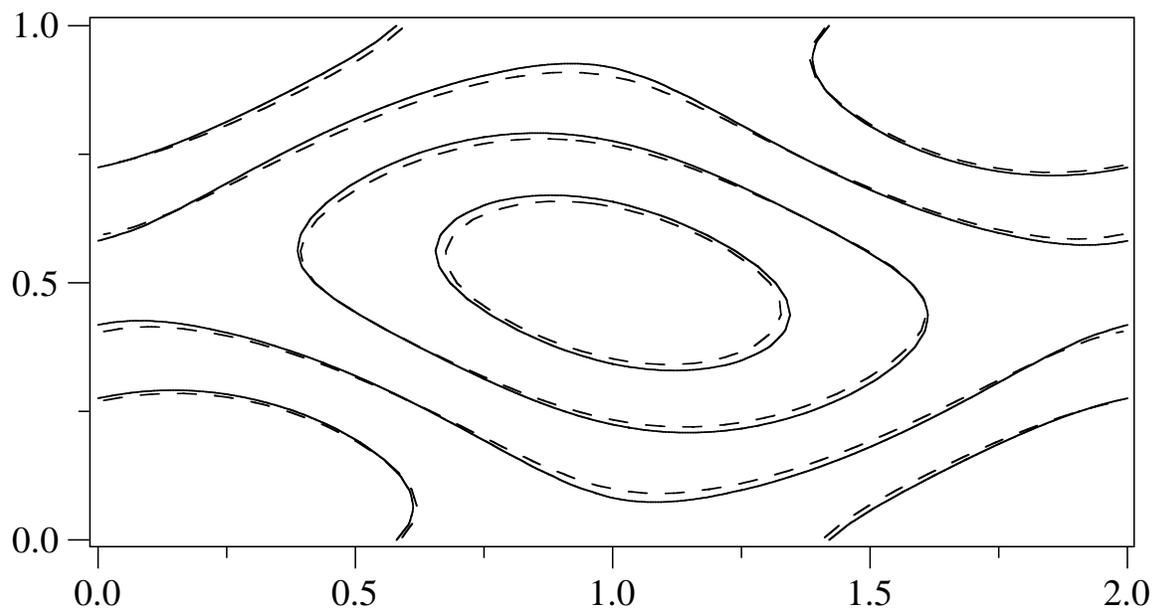
\begin{figure}
\label{fig2}
\caption{ Stationary state contour plot of the stream function for $R =
15$, $a_r
= 2$ and $\theta_0 = 0$. The full and dashed lines correspond to theoretical
prediction  (eq. \ref{streamtheory}) and numerical results, respectively.  The
discrepancy remains below $5 \%$. }

\end{figure}

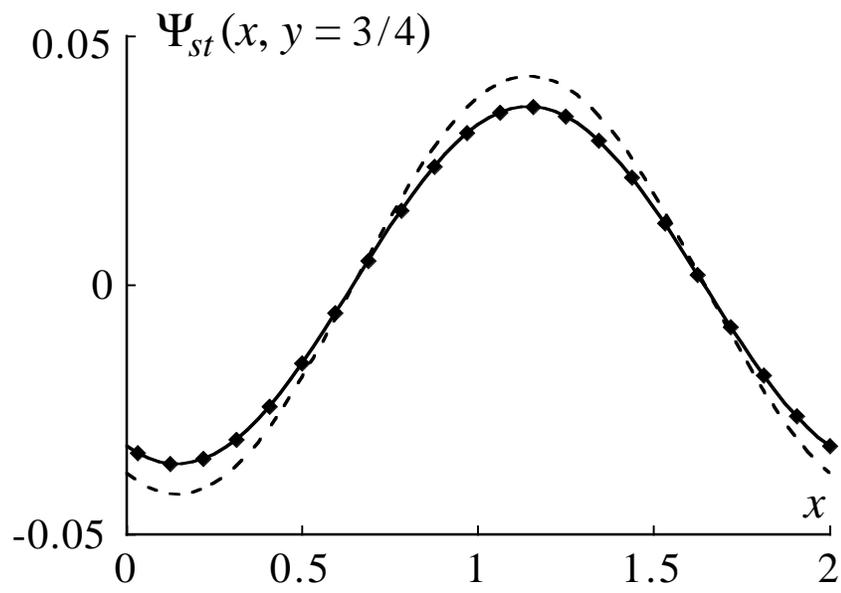
\begin{figure}
\label{fig3}
\caption{ Horizontal profile of the stationary state
stream function, with $y = 3/4$, as a function of the vertical coordinate
$x$ for $R = 13$, $a_r
= 2$ and $\theta_0 = 0$.  The full and dashed lines represent theoretical
predictions obtained by using an estimation of the critical Reynolds number
based
on 5-modes (eq. \ref{Rc5}) and 3-modes (eq. \ref{Rc}) approximation theories,
respectively.  The diamonds correspond to numerical results.}
\end{figure}

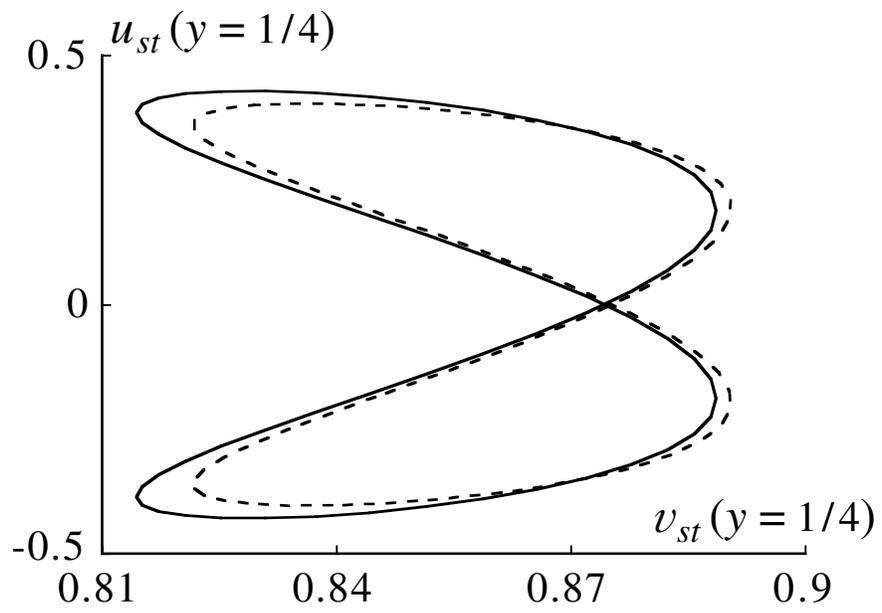
\begin{figure}
\label{fig4}
\caption{Vertical versus horizontal components of the stationary state velocity
field with  $y = 3/4$.  The full line corresponds to theoretical
predictions, as given by eqs. (\ref{ustationar}) and (\ref{vstationar}),
whereas
the dashed line is obtained by solving numerically the
compressible nonlinear hydrodynamic equations.  The parameters are  $R = 15$,
$a_r = 2$, $\theta_0 = 0$ and
$\varepsilon = 10^{-2}$.  The discrepancy is about
$5 \%$.}
\end{figure}

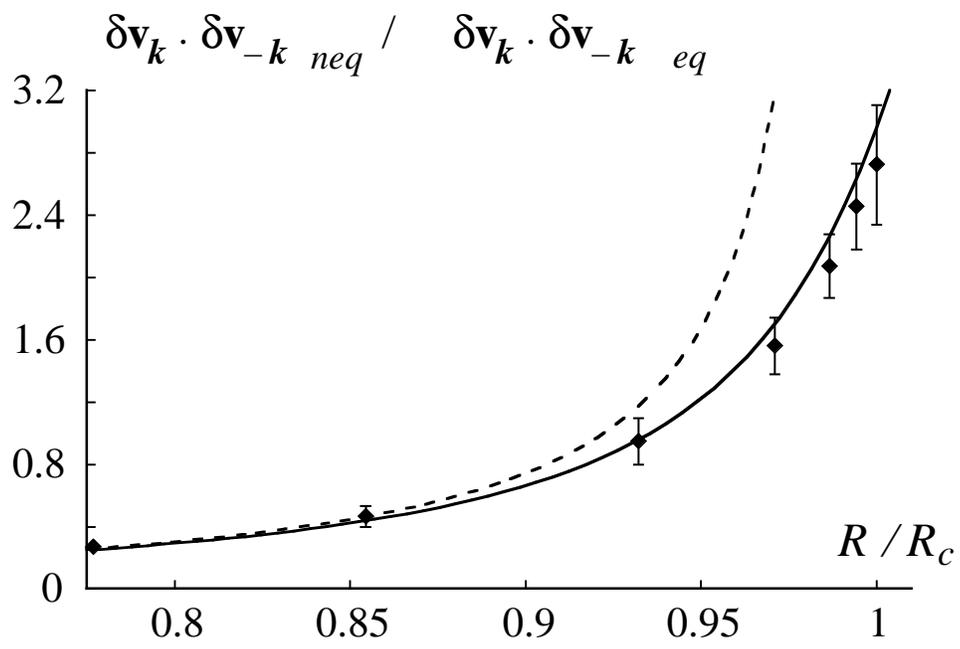
\begin{figure}
\label{fig5}
\caption{Fourier transform of the nonequilibrium part of the static
velocity auto-correlation
function, normalized by the corresponding equilibrium part, as a function
of $R/R_c$.
The solid and dashed curves represent the complete and the
linearized solutions, eqs. (\ref{delvNL}) and (\ref{delvLin}),
respectively, whereas the
diamonds correspond to numerical results obtained by the simulation of
nonlinear
compressible fluctuating hydrodynamic equations.  The parameters are $a_r =
2$, $\varepsilon =
10^{-2}$ and
 ${\cal A} = 10^{-3} / 256$. The
estimated statistical error is about $13 \%$ for the last data point.}
\end{figure}

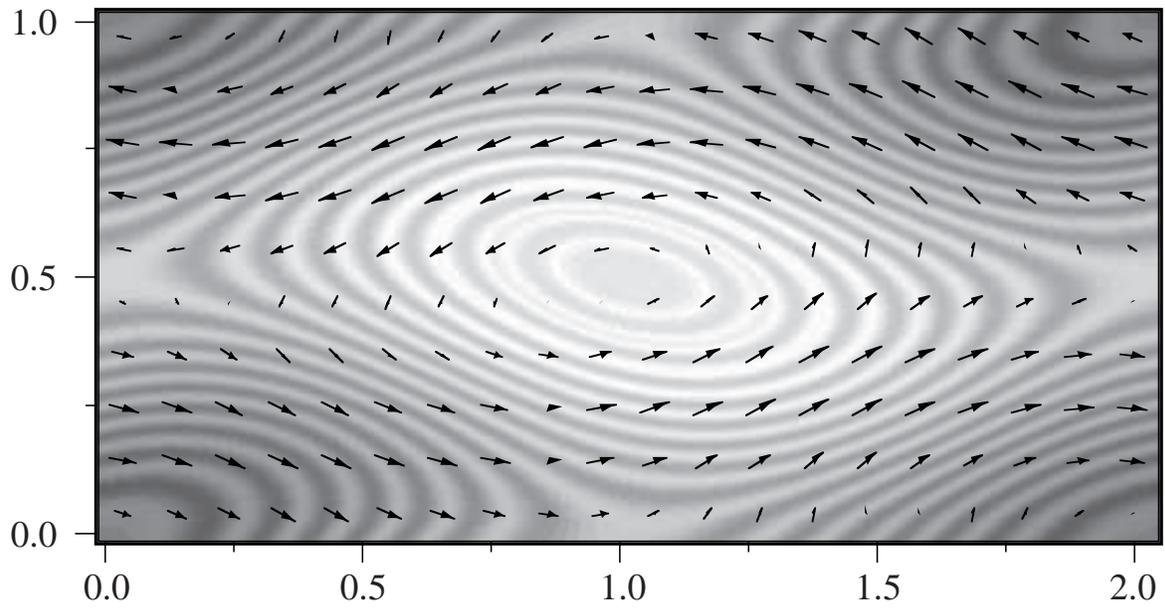
\end{document}